\documentclass[fleqn,twoside,epsfig]{article}
\usepackage{espcrc2}
\usepackage{amssymb}

\usepackage{graphicx}
\usepackage[figuresright]{rotating}

\newcommand{\Fig}[1]{Fig.~\ref{#1}}
\newcommand{\Eq}[1]{Eq.~(\ref{#1})}
\newcommand{\be}{\begin{equation}}
\newcommand{\ee}{\end{equation}}
\newcommand{\bea}{\begin{eqnarray}}
\newcommand{\eea}{\end{eqnarray}}

\newcommand{\rf}[4]{{\em {#1}} {\bf #2}, #3 (#4)}

\newcommand{\slh}{\!\!\!\slash}

\newcommand{\cm}{Commun.\ Math.\ Phys.}

\title{ 
\vspace{-0.6cm}
Nonperturbative renormalisation of composite operators with 
overlap quarks\thanks{Presented by J. B. Zhang}}
\author{J.\ B.\ Zhang\address[CSSM]{CSSM Lattice Collaboration,\\
Special Research Center for the Subatomic Structure of
Matter (CSSM) and Department of Physics,
University of Adelaide 5005, Australia},
D.\ B.\ Leinweber\addressmark[CSSM],
K.\ F.\ Liu\address[UK]{Department of Physics and Astronomy, University of Kentucky, Lexington, KY 40506}, 
A.\ G.\ Williams\addressmark[CSSM]}
\date{\today}

\begin{document}

\begin{abstract}
We compute non-perturbatively the renormalisation constants of composite operators 
on a $16^3 \times 32 $ lattice with lattice spacing $a$ = 0.093 fm  
for the overlap fermion action by using the regularisation independent (RI) scheme. 
The quenched gauge configurations are generated by
tadpole improved plaquette plus rectangle action. We test the perturbative continuum 
relation $Z_A = Z_V$
and $ Z_S=Z_P$ and find that they agree well above $\mu$ = 1.6 GeV.
We also perform a Renormalisation Group analysis at the
next-to-next-to-leading order and convert the renormalisation constants to the
$\overline{MS}$ scheme.
 
\end{abstract}


\maketitle

\section{INTRODUCTION}

Lattice QCD is a unique tool with which we can compute physical observables 
non-perturbatively from first principles.
Renormalisation of lattice operators is an essential ingredient needed to
deduce physical results from numerical simulations.
In this paper we study the renormalisation properties of composite bilinear
operators with the overlap quark action.

In principle, the renormalisation of a quark bilinear can be computed
by lattice perturbation theory. However, lattice perturbation theory converges
slowly and the higher-order corrections may not be small, thus introducing a 
large uncertainty in the calculation of the renormalised matrix elements
in some continuum scheme.
To overcome these difficulties, Martinelli {\it et al.}
\cite{NPM} have proposed a promising
non-perturbative renormalisation procedure.
The procedure allows a full non-perturbative computation of the matrix elements
of composite operators in the Regularisation Independent (RI) scheme
\cite{NPM,eps'/eps}. The matching between the RI scheme and $\overline{MS}$, which is
intrinsically perturbative, is computed using only continuum perturbation
theory and has been carried out to higher orders, which is well behaved.
This method has been shown to be quite successful in reproducing results
obtained by other methods, such as using chiral Ward Identities (WI) \cite{Ward}.
The method
has also been successfully applied to determine renormalisation
coefficients for various operators using the Wilson
\cite{noi_mq,DS99,DS992,becirevic98} and
staggered actions \cite{JLQCD1}, domain-wall fermions~\cite{Tblum1}, as well as
the quark mass renormalisation constant for overlap fermions~\cite{GHR}.
The purpose of the current work is to study the
application of this non-perturbative renormalisation procedure to the 
renormalisation of the quark field and
the flavour non-singlet fermion bilinear operators in the case of overlap fermions.


We shall use Neuberger's overlap fermions~\cite{neub2} which have lattice
chiral symmetry at finite cut-off. As a result, many chiral-symmetry relations
are still valid~\cite{neub1,luscher} and the quark propagator~\cite{liu02} 
preserves the same structure as in the continuum.
The use of the
overlap action entails many theoretical advantages~\cite{neub3,kent1}, such
as no additive mass renormalisation, no ${\mathcal O(a)}$ error, and no mixing among 
operators of different chirality. The latter is very helpful for computing weak 
matrix elements.

\section{NON-PERTURBATIVE RENORMALIZATION METHOD}
\label{sec:NPRM}

  In this section we review the nonperturbative renormalisation method of Ref.~\cite{NPM},
which we use to compute the renormalisation constants of quark bilinears in this paper.
The method imposes renormalisation conditions non-perturbatively, directly on quark and 
gluon Green's functions, in a fixed gauge, for example the Landau gauge. 

Let $S\left(x,0\right)$ denotes the quark propagator on a gauge-fixed configuration
from a source $0$ to all space-time points $x$. 
The momentum space propagator is
defined as the discrete Fourier transform over the sink positions
\begin{equation}
S\left(p,0\right) =
\sum_x \exp \left( -i p\cdot x \right) S \left( x, 0 \right) \, .
\end{equation}

We use periodic boundary conditions in spatial directions and an 
anti-periodic boundary condition in time direction. 
The dimensionless lattice momenta
are
\begin{equation}
p_i=\frac{2\pi}{N_s}(n_i-\frac{N_S}{2}) \, ,~~
p_t=\frac{2\pi}{N_t}(n_t-\frac{1}{2}-\frac{N_t}{2})
\label{eq:pi}
\end{equation}
for an $N_s^3 \times N_t$ lattice, where $n_i$ ($n_t$) may in principle
lie in the range $ 0 \rightarrow N_s (N_t)$. In practice, however, 
only a subset of this range is used. In this paper the momentum
range was restricted to those momenta for which $n_i = 0,1,2$ 
and $n_t = 1,2,3,4 $.

We also define the square of the absolute
momentum, 
\begin{equation}
\left( pa\right)^2 = a^2 \sum_{\mu = 0}^{3} p_\mu \, p_\mu
\, .
\end{equation}

\subsection{ Three Point Function}

Consider the two-fermion operators
\begin{equation}
O_{\Gamma}(x) = \bar{\psi}(x) \Gamma \psi \, ,
\end{equation}
where $\Gamma$ is the Dirac gamma matrix
\begin{equation}
\Gamma \in \left\{ 1 , \gamma_\mu , \gamma_5, \gamma_\mu
\gamma_5, \sigma_{\mu \nu} \right\} \, ,
\end{equation}
and the corresponding notation will be \{S, V, P, A, T\} respectively.
The three point function with the operator insertion at position $0$ and
the propagator from $y$ to $0$ and then from $0$ to $x$ is given by
\begin{eqnarray}
G_O (x, 0, y) & = &\langle \psi(x) O_{\Gamma}(0) \bar{\psi}(y)\rangle \nonumber \\
 &= &\langle S(x,0) \Gamma S(0,y) \rangle \, ,
\end{eqnarray}
where $S(0,y)$ is the quark propagator from $y$ to $0$. It is the inverse of the
Dirac operator. 

The Fourier transform of the three-point function is given by
\begin{eqnarray}
G_O(pa,qa) &\equiv& \int d^4 x d^4 y e^{-i (p \cdot x - q \cdot y)} G_O (x, 0, y) \nonumber \\
  & = & \langle S(p,0) \Gamma  \left(\gamma_5 S^{\dagger}(q,0)\gamma_5\right)\rangle \, .
\end{eqnarray}
From this, one can define the amputated three-point function
\begin{equation}
\Lambda_O (pa, qa) = S(pa)^{-1} G_O(pa,qa) S(qa)^{-1} \, ,
\end{equation}
\label{lambdao}
where
\begin{equation}
S(pa) = \langle S(p,0) \rangle \, ,
\end{equation}
which is translational invariant and a $12 \times 12$ matrix.

   Finally, we define a projected vertex function
\begin{equation}
\Gamma_O(pa) = \frac{1}{\mathrm{Tr}(\hat{P}_O^2)} \mathrm{Tr} \left(\Lambda_O (pa, pa) \hat{P}_O\right)\, ,
\label{eqvertex}
\end{equation}
where $\hat{P}_O = \Gamma$ is the corresponding projection operator.

\subsection{RI/MOM Renormalisation Condition}


The renormalised operator $O(\mu)$ at scale $\mu$
is related to the bare operator
\begin{equation}
O(\mu) = Z_O(\mu a, g(a)) O(a) \, ,
\end{equation}
and the renormalisation condition is imposed on the vertex function $\Gamma_O(pa)$
at a scale $p^2 = \mu^2$,
\begin{equation}    \label{eq:ren_con}
\Gamma_{O, \rm{ren}}(pa)|_{p^2 = \mu^2} = \frac{Z_O}{Z_{\psi}}
\Gamma_O(pa)|_{p^2 = \mu^2} = 1 \, ,
\end{equation}
to make it agree with the tree-level value of unity. Here $Z_{\psi}$ is the
field or wavefunction renormalisation
\begin{equation}
\psi_{\rm{ren}} = Z_{\psi}^{1/2} \psi \, .
\end{equation}

   In order to obtain the renormalisation constant $Z_O$ for the operator $O$, one needs
to know $Z_{\psi}$ first. It has been suggested that $Z_{\psi}$ be obtained from the
renormalisation of vector or axial-vector currents. For example, if one uses
the conserved vector current, then one expects $Z_{V^C} = 1$. Therefore, from the
renormalisation condition \Eq{eq:ren_con},
one obtains,
\begin{equation}
Z_{\psi} = \frac{1}{48} {\mathrm{Tr}}\left(\Lambda_{V_{\mu}^C}(pa) \gamma_{\mu}\right)|_{p^2 = \mu^2} \, .
\end{equation}

  However, in this work, we will obtain $Z_{\psi}$ directly from the quark propagator. It
can be defined from the Ward Identity (WI) as \cite{NPM}
\begin{equation}
Z_\psi=-i
\frac{1}{12} {\mathrm{Tr}} \left[\frac{\partial S(pa)^{-1}}{\partial \slh{p}}\right]\left.
\right|_{p^2=\mu^2} \; .
\label{eq:Z_q_WI}
\end{equation}
To avoid derivatives with respect to a discrete variable, we have used
\begin{equation}
Z'_\psi=\left.-i\frac{1}{12}\frac{{\mathrm{Tr}} \sum_{\mu=1,4}
\gamma_\mu (p_\mu a)S(pa)^{-1}}
{4\sum_{\mu=1,4}(p_\mu a)^2}\right|_{p^2=\mu^2}\; ,
\label{eq:Z_q'_WI}
\end{equation}
which, in the Landau Gauge, differs from $Z_\psi$ by a finite term of order
$\alpha_s^2$ \cite{Franco:1998bm}, and the difference is less than 1\% at the 
typical scale $\mu~ \sim ~ $ 2 GeV. It has been pointed out in Ref.~\cite{Tblum1} that 
the systematic error due to the definition of lattice momentum $p$ is much larger than 1\%. We will not
distinguish $Z'_\psi$ and $Z_\psi$ later on.


\section{NUMERICAL RESULTS}
\label{sec:numerical}

 We work on a $16^3\times{32}$ lattice with lattice spacing, $a$=0.093 fm.
The gauge configurations are created using a tadpole improved plaquette plus
rectangle (L\"{u}scher-Weisz~\cite{Lus85}) gauge action through the
pseudo-heat-bath algorithm. 
A total of 50 configurations are considered.
The lattice parameters are summarised in Table~\ref{simultab}.  The lattice
spacing is determined from the static quark potential with a string
tension $\sqrt{\sigma}=440$~MeV~\cite{zanotti}.

\vspace*{-0.7cm}
\begin{table}[ht]
\caption{\label{simultab}Lattice parameters.}
\begin{tabular}{ccccc}
\hline
Action &Volume  & $N_{\rm{Samp}}$ &$\beta$ &$a$ (fm) \\
\hline
Improved       & $16^3\times{32}$ & 500 & 4.80 & 0.093  \\
\hline
\end{tabular}
\end{table}
\vspace*{-0.7cm}

The gauge field configurations are gauge fixed to the Landau gauge using a
Conjugate Gradient Fourier Acceleration~\cite{cm} algorithm with an accuracy
of $\theta\equiv\sum\left|\partial_{\mu}A_{\mu}(x)\right|^{2}<10^{-12}$.
We use an improved gauge-fixing scheme~\cite{bowman2} to minimise
gauge-fixing discretisation errors. 

The massive overlap Dirac operator~\cite{neub2} is defined so that at
tree-level there is no mass or wavefunction renormalisation~\cite{ddh02},
\begin{equation}
D(m_0) = \rho + \frac{m_0}{2} + (\rho - \frac{m_0}{2} ) \gamma_5 \epsilon (H) \, .
\end{equation}
Here $\epsilon (H) = H /\sqrt{H^2}$ is the matrix sign function and $H$ is
taken to be the hermitian Wilson-Dirac operator, {\it i.e.}, $H = \gamma_5 D_{\rm
W}$.  Here $D_{\rm W}$ is the usual Wilson fermion operator, but with a
negative mass parameter $- \rho = 1/2\kappa -4$ in which $\kappa_c < \kappa <
0.25$. 

Our numerical calculation begins with an evaluation of the
inverse of $D(m_0)$ 
for each gauge configuration in the ensemble. We use a 14th order Zolotarev approximation~\cite{zolo}
to the sign matrix function $\epsilon(H_w)$, and in the selected window of $x \in [0.031,2.5] $ of
$\epsilon(x)$, the approximation  is better than $3.3 \times 10^{-10}$~\cite{kent3}. 
In the calculations, $\kappa=0.19163$  is used, which gives $\rho a= 1.391$.
We calculate 15 quark masses by using a shifted version of
a Conjugate Gradient solver.
The  bare quark masses $m_0 a$  are chosen to be 
$0.0250$, $0.0278$, $0.0334$, $0.0389$, $0.0445$, $0.0501$,
$0.0584$, $0.0668$, $0.0834$, $0.1001$, $0.1252$, $0.1669$, $0.2086$, $0.2503$, $0.2921$.
 The $m_0$ values in physical units are
 $53$, $59$, $71$, $83$, $94$, $106$, $124$, $142$, $177$, $212$, $266$,
 $354$, $442$, $531$, and $620$~MeV respectively.

\subsection{Axial and vector currents}

\begin{figure}[tb]
{\small
\resizebox*{\columnwidth}{4.8cm}{\rotatebox{90}{\includegraphics{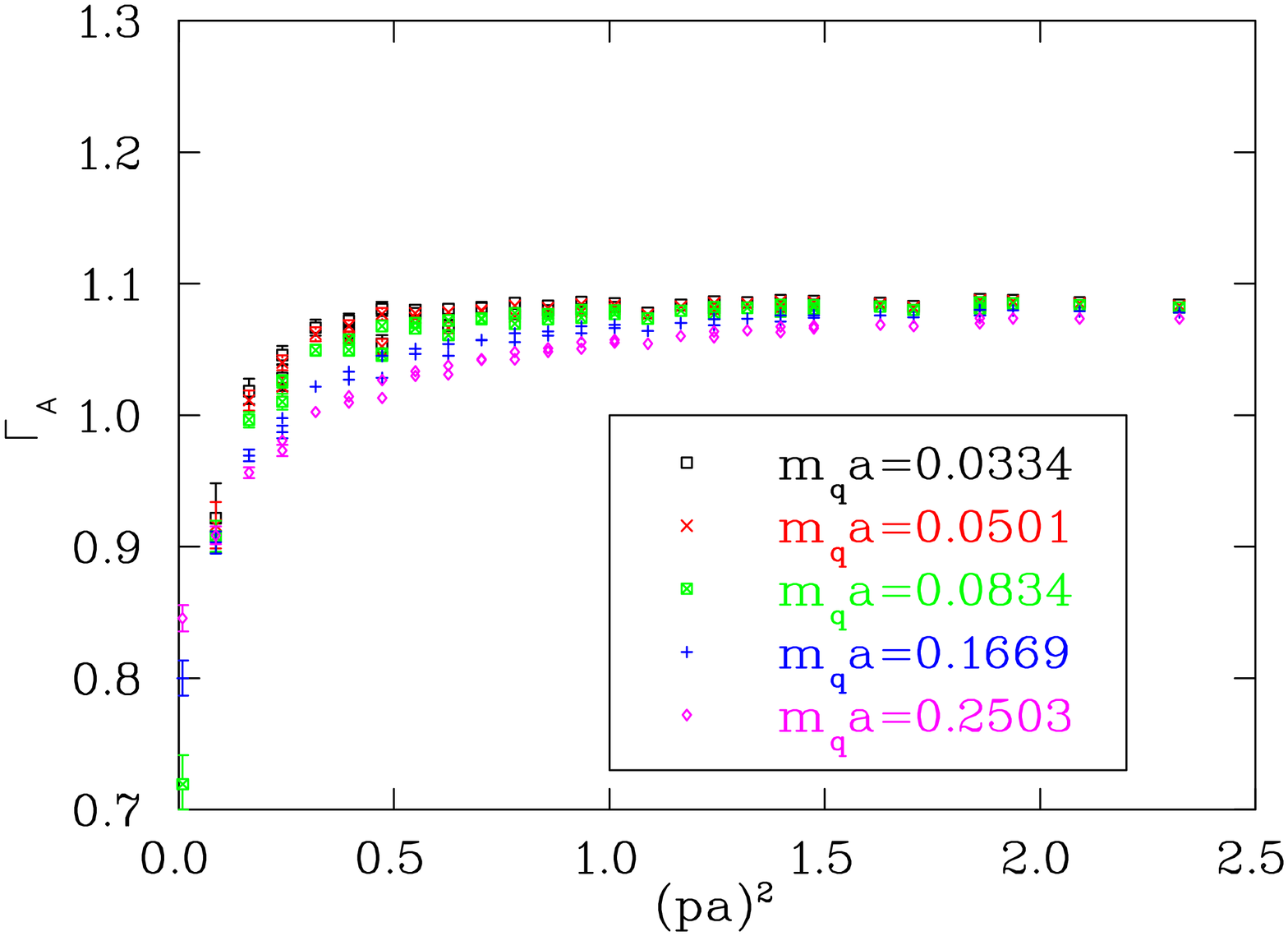}}}
\vspace*{-0.7cm}
\resizebox*{\columnwidth}{4.8cm}{\rotatebox{90}{\includegraphics{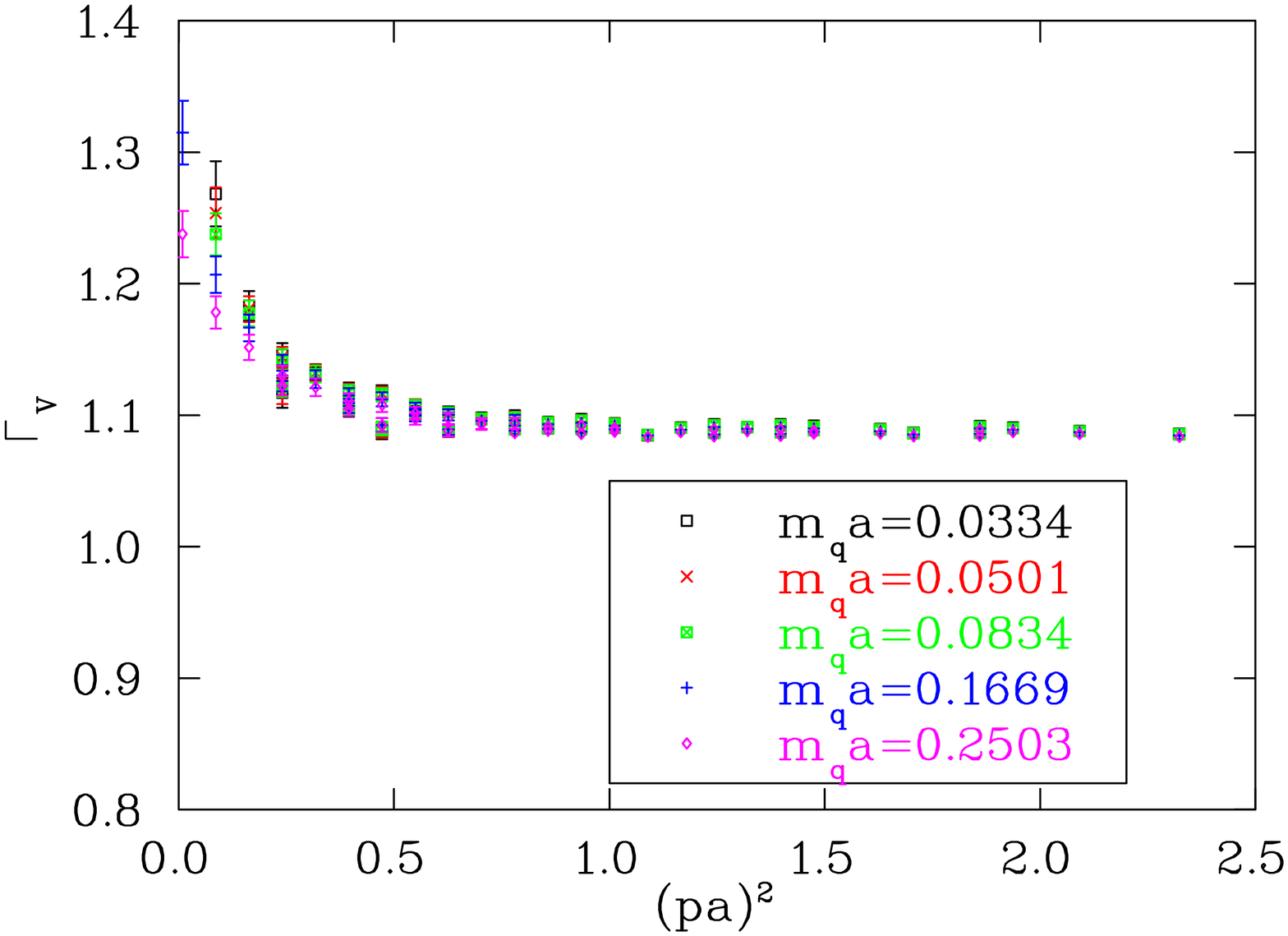}}}
}
\vspace*{-0.7cm}
\caption{\label{figav}{
\small The projected vertex function $\Gamma$ defined in Eq.~(\ref{eqvertex})
for vector and axial vector currents with different bare quark masses. 
By the renormalisation condition (Eq.(~\ref{eq:ren_con})), 
they are the ratio of renormalisation constant
$Z_{\psi}/Z_A$ and $Z_{\psi}/Z_V$ respectively.}}
\vspace*{-0.7cm}
\end{figure}

\begin{figure}[tb]
{\small
\resizebox*{\columnwidth}{4.8cm}{\rotatebox{90}{\includegraphics{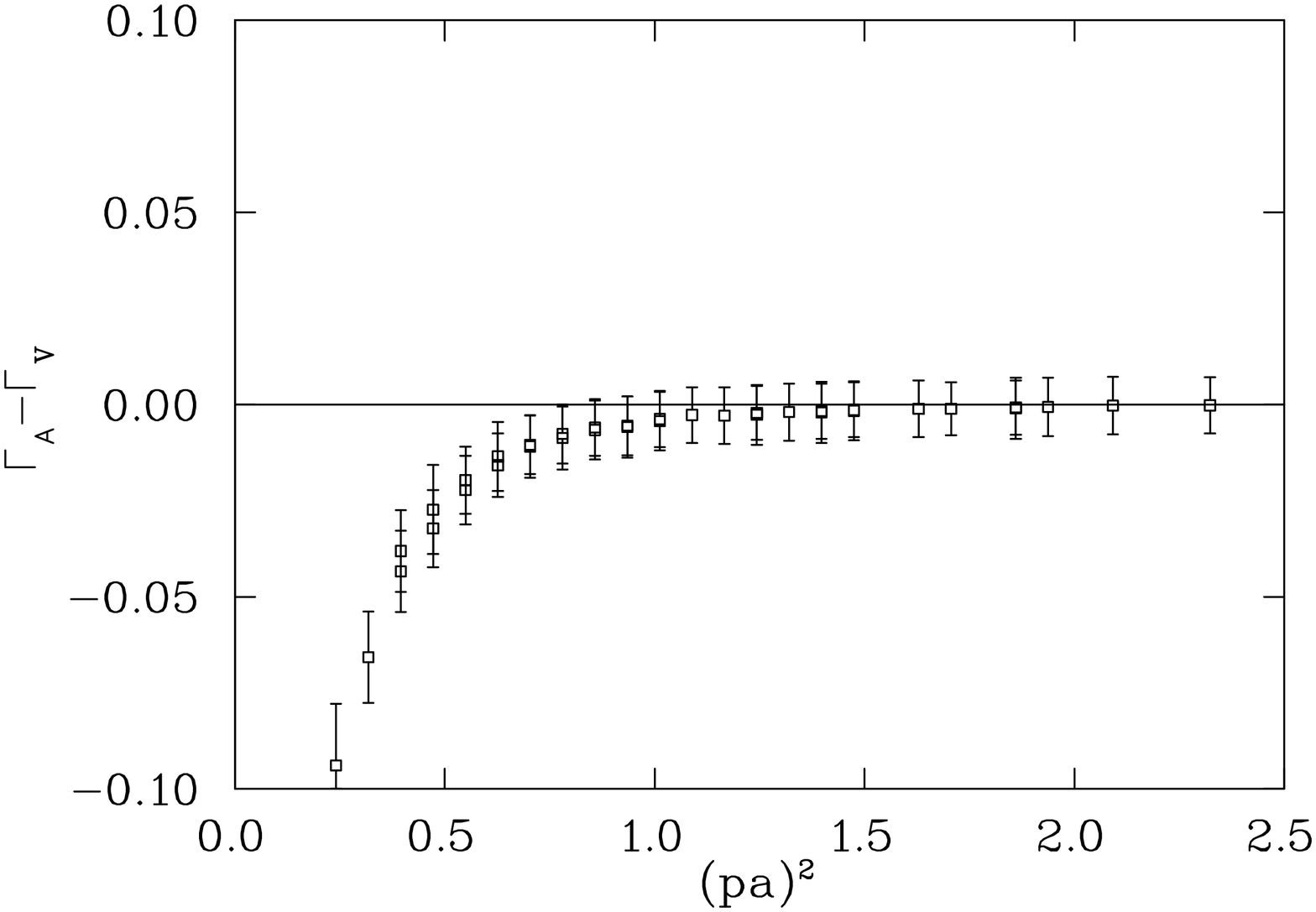}}}
\vspace*{-0.7cm}
\resizebox*{\columnwidth}{4.8cm}{\rotatebox{90}{\includegraphics{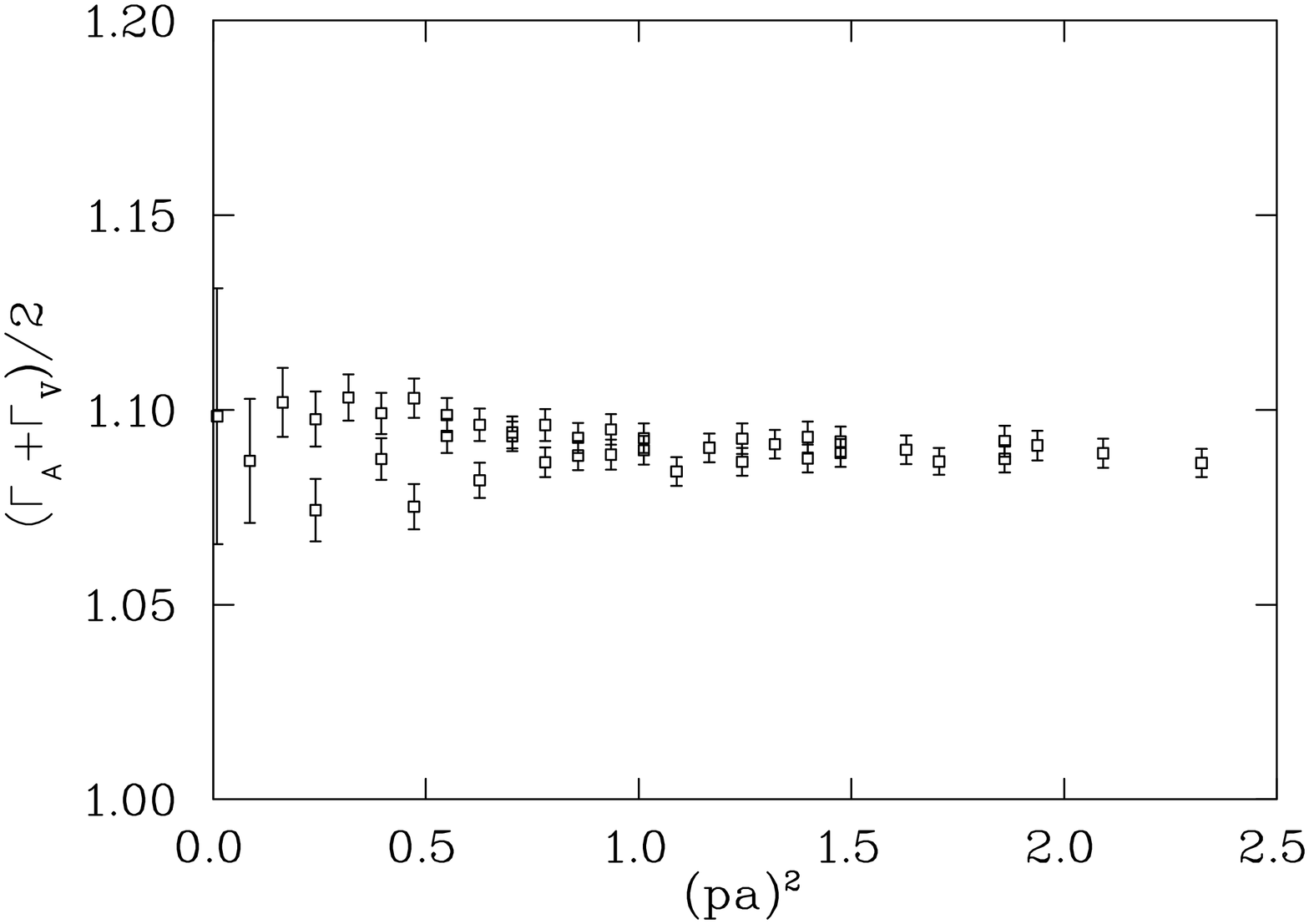}}}
}
\vspace*{-0.7cm}
\caption{\label{Checkva}{\small
Checking $Z_A$ = $Z_V$, the upper plot is ($\Gamma_A-\Gamma_V$)
versus $(pa)^2$, where we linear extrapolate to the chiral limit ($m_0=0$).
The lower plot is $\frac{1}{2}$ ($\Gamma_A+\Gamma_V$)
versus $(pa)^2$. The relation $Z_A$ = $Z_V$ is valid at moderate to large $(pa)^2$ .
}}
\vspace*{-0.7cm}
\end{figure}

Let us consider first the vector and axial vector currents. Since each obeys a
chiral WI \cite{Bochicchio} their renormalisation constants
are finite. In Fig.~\ref{figav}, we show $\Gamma_A$ and $\Gamma_V$, calculated
in the RI scheme for different bare quark masses as a function of the 
lattice momentum $(pa)^2$. We find they are weakly dependent on the mass, 
and almost scale independent for $ (pa)^2~ \ge ~0.5 $.

For the overlap fermion, which satisfies the Ginsparg-Wilson~\cite{GW} 
relation and preserves 
chiral symmetry on finite lattice. In  Ref.~\cite{vicari} it has been proved that the renormalisation constants
for naive vector and axial vector currents satisfy
\be 
Z_A = Z_V \, .  
\ee

In Fig.~\ref{Checkva} we show the quantities $\Gamma_A - \Gamma_V $ and
and $\frac{1}{2}(\Gamma_A + \Gamma_V) $, after extrapolating to the
chiral limit. We can see from upper part of the figure that 
there is no signal of 
effects from chiral symmetry breaking, since $\Gamma_A- \Gamma_V$
is tending to zero at moderate and high momenta. 
The effects of spontaneous chiral symmetry breaking are visible at low
momenta and they are damped at higher momenta.
At the momenta of interest, there also seems to be
no significant splitting due to non-perturbative effects with the
difference between $\Gamma_A$ and $\Gamma_V$ being less than
$1\%$  at $(ap)^2=0.7$ and smaller for momenta above
this. In the lower part of Fig.~\ref{Checkva}, we plot $\frac{1}{2}(\Gamma_A + \Gamma_V) $
against $(ap)^2$, and it can be used for the extraction of 
both $Z_A/Z_\psi$ and $Z_V/Z_\psi$ to increase the statistical accuracy.

\subsection{Pseudoscalar and scalar densities}

\begin{figure}[t]
{\small
\resizebox*{\columnwidth}{4.8cm}{\rotatebox{90}{\includegraphics{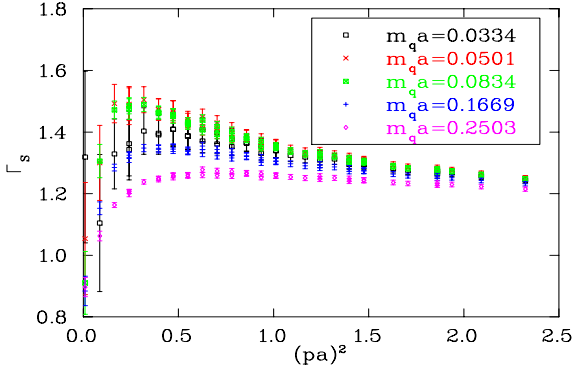}}}
}
\vspace*{-1.2cm}
\caption{\label{figgammas}{\small
$\Gamma_S$ versus $(pa)^2$ with different masses. Here
one can see the strong mass dependence and non-monotonic behaviour.
}}
\vspace*{-0.8cm}
\end{figure}

\begin{figure}[tb]
{\small
\resizebox*{\columnwidth}{4.8cm}{\rotatebox{90}{\includegraphics{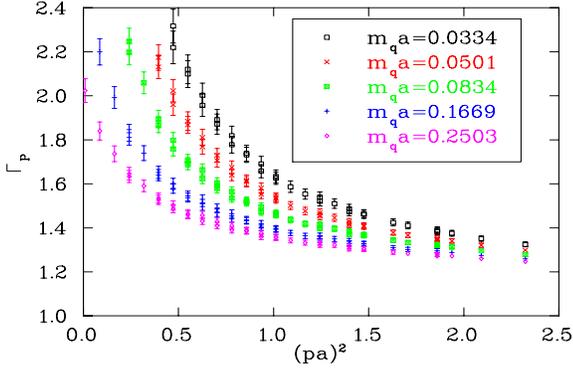}}}
}
\vspace*{-1.2cm}
\caption{\label{figgammap}{\small
$\Gamma_P$ versus $(pa)^2$ with different masses. Here also
we have strong mass dependence and pole behaviour.
}}
\vspace*{-0.8cm}
\end{figure}

\begin{figure}[t]
{\small
\resizebox*{\columnwidth}{4.8cm}{\rotatebox{90}{\includegraphics{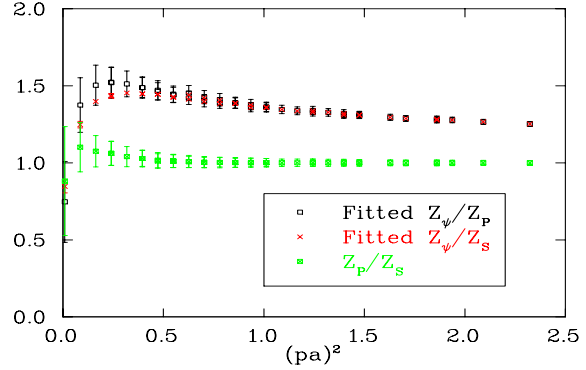}}}
}
\vspace*{-1.2cm}
\caption{\label{checksp}{\small
Checking of $Z_P$ = $Z_S$ after mass pole subtraction.
}}
\vspace*{-0.6cm}
\end{figure}

\begin{figure}[tb]
{\small
\resizebox*{\columnwidth}{4.8cm}{\rotatebox{90}{\includegraphics{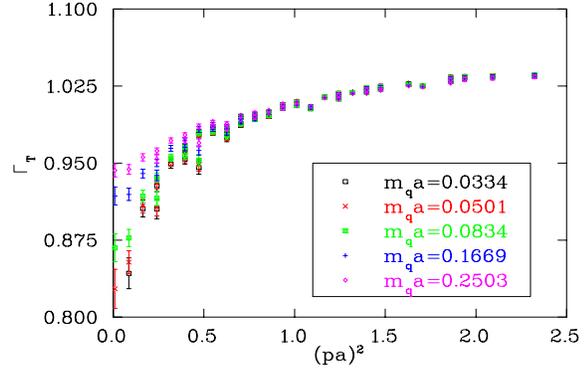}}}
}
\vspace*{-1.2cm}
\caption{\label{figgammt}{\small
$\Gamma_T$ versus $(pa)^2$ with different masses.
It shows no significant mass dependence for moderate to large $(pa)^2$.
}}
\vspace*{-0.8cm}
\end{figure}

The pseudoscalar and scalar densities differ from the axial and vector
currents in that their renormalisation is not 
scale independent. 
For overlap fermions with chiral symmetry the
RI scheme preserves the well known $\overline{MS}$ relations
\be
Z_S = Z_P   ~~~{\rm and} ~~
Z_m = \frac{1}{Z_S} \, ,
\ee
and the
quantities $Z_S/Z_P$, $Z_SZ_m$ and $Z_m Z_P$ are expected
to be scale independent.

Fig.~\ref{figgammas} and Fig.~\ref{figgammap} show $\Gamma_S$ 
and $\Gamma_P$ versus $(pa)^2$ with different masses. 
Unlike in the case of the vector and axial vector currents, 
they are strongly dependent on the quark mass $m_0$ for light quarks.
When we want to extract the chiral limit value, 
we cannot use a simple linear or quadratic
fit. In Fig.~\ref{figgammas} and Fig.~\ref{figgammap}, we can clearly 
see the pole effect in the small $(pa)^2$ region.

As discussed in detail in Ref.~\cite{Tblum1}, due to
the contribution from the zero-modes, we may fit the $\Gamma_S$ to the form
\be
\Gamma_S
 =
\frac{c_{1,S}}{(am_0)^2} + c_{2,S} +
c_{3,S} (am_0)^2 
\ee
at a fixed momentum 
with $Z_\psi/Z_S$ being given by $c_{2,S}$. The fitted value of $ c_{2,S}$ is 
plotted in Fig.~\ref{checksp} as $Z_\psi/Z_S$. We find that the  
coefficient of the pole term, $c_{1,S}$  is very small, 
between $10^{-3}$ to $10^{-4}$, this means that the pole term 
only provides a large contribution at very small $m_0$, i.e., near the chiral limit.

 For $\Gamma_P$, we fit to the form
\be
\Gamma_{P,{\rm latt}}
 =
\frac{c_{1,P}}{(am_0)^2}
+
\frac{c_{2,P}}{(am_0)}
+
c_{3,P} +
c_{4,P} ( am_0 )^2 \, ,
\label{eq79}
\ee
with $c_{3,P}$ being equal to $Z_\psi/Z_P$. The quadratic mass pole is due to
zero-mode effects in $\langle\overline{q}q\rangle$.  The fitted value of $ c_{3,P}$ 
is plotted in Fig.~\ref{checksp} as $Z_\psi/Z_P$.

As the  evidence for the above fitting form, we  consider
the resulting values for $Z_\psi/Z_S$ and $Z_\psi/Z_P$. Fig.~\ref{checksp} shows a
comparison between the extracted values of these two quantities. As chiral
symmetry would predict for $Z_\psi/Z_S$ and $Z_\psi/Z_P$, the two quantities
coincide at moderate and large momenta.This provides an excellent test of both the
fitting method to extract the poles and the chiral properties of 
overlap fermions.

\subsection{The Tensor current}

In Fig.~\ref{figgammt} we show $\Gamma_T$
versus $(pa)^2$ with different masses.  We can see that at moderate and large
$(pa)^2$, $\Gamma_T$ is not sensitive to the quark masses. 
The chiral limit value
is obtained by a linear fit in quark mass as for the cases of vector and axial vector currents,
and the results will be presented in the next subsection.

\subsection{Running of the renormalisation constants}

\begin{figure}[tb]
{\small
\resizebox*{\columnwidth}{4.8cm}{\rotatebox{90}{\includegraphics{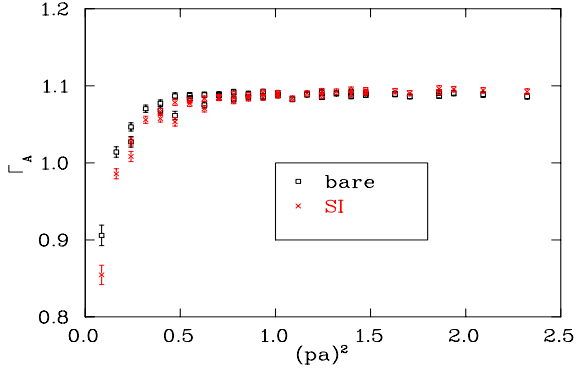}}}
}
\vspace*{-1.0cm}
\caption{\label{figvasi}{\small
$\Gamma_A$ versus $(pa)^2$ in the chiral limit 
before (labelled as ``bare'') and after (labelled as ``SI'') three
loop perturbative running, such that they coincide at $(pa)^2=1$. The 
latter is almost $(pa)^2$ independent after $(pa)^2 > 0.6$. 
The slope versus $(pa)^2$ is about 0.01 
and is interpreted as an $O(a^2)$ effect.  }}
\vspace*{-0.8cm}
\end{figure}

\begin{figure}[tb]
{\small
\resizebox*{\columnwidth}{4.8cm}{\rotatebox{90}{\includegraphics{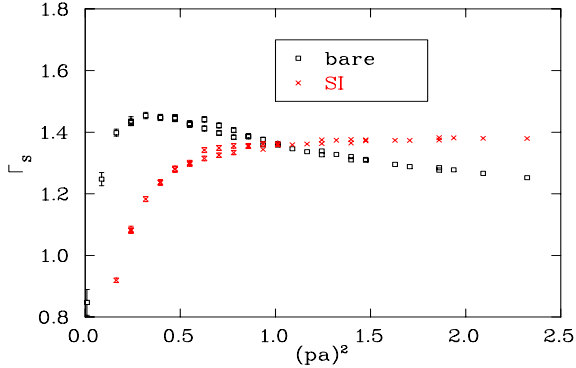}}}
}
\vspace*{-1.0cm}
\caption{\label{figvssi}{\small
The same as Fig.~(\ref{figvasi}). Here is $\Gamma_S$ versus $(pa)^2$. 
The slope of SI versus $(pa)^2$ is about 0.02.
}}
\vspace*{-0.8cm}
\end{figure}
 
\begin{figure}[tb]
{\small
\resizebox*{\columnwidth}{4.8cm}{\rotatebox{90}{\includegraphics{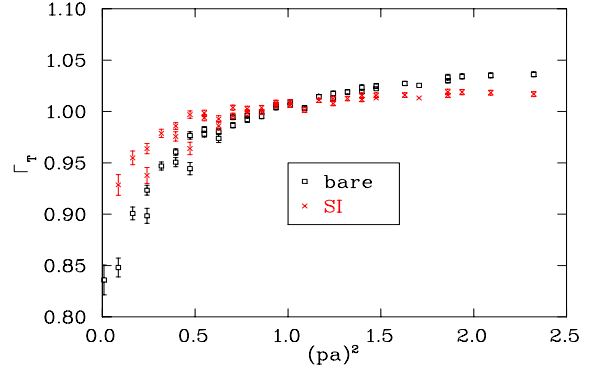}}}
}
\vspace*{-1.0cm}
\caption{\label{figvtsi}{\small
$\Gamma_T$ versus $(pa)^2$ in the chiral limit before and after two
loop perturbative running. The latter  is almost
$(pa)^2$ independent after $(pa)^2 > 0.6$, 
the slope versus $(pa)^2$ is about 0.01.
}}
\vspace*{-0.8cm}
\end{figure}

The renormalised operators are  defined as
\begin{equation}
Z_{O} O_{\rm{bare}} = O_{\rm{ren}} \, .
\end{equation}
Requiring the bare operator to be independent of the renormalisation
scale $\mu^2$ gives the RG equation,
\be
\mu^2 \frac{d}{d \mu^2} O_{\rm{ren}}
=\frac{1}{Z_O} \mu^2 \frac{d Z_{O}}{d \mu^2}  O_{\rm{ren}} 
=
-\frac{\gamma_O}{2} O_{\rm{ren}} \, .
\ee
The solution of $Z_O(\mu^2)$ can be written in the following
form
\begin{equation}   \label{eq:evo}
Z_O\left(\mu^2\right) =
\frac{
C_O\left({\mu}^2\right)
}{
C_O\left({\mu'}^2
\right)
} Z_O\left({\mu'}^2\right) \, .
\end{equation}
The coefficient $ C_O\left({\mu}^2\right)$ can be found in 
Appendix B of Ref.~\cite{Tblum1} and references within. 

In this work, the value of $\alpha_s$ was calculated at three loops using
a lattice value of $\Lambda_{\rm{QCD}}$
taken from Ref.~\cite{Capitani:1998mq}
as
$ \Lambda_{\rm{QCD}} = 238 \pm 19 \mathrm{MeV} $ . 

Both $Z_A$ and $Z_V$ should be scale independent, but this is not the case for
$Z_\psi$. Fig.~\ref{figvasi} shows both $\Gamma_A$ and the 
scale invariant (SI) quantity
(the data after removing the renormalisation group running) 
calculated as described above:
\be
\Gamma_A^{SI}\left((ap)^2\right) = \Gamma_A\left((ap)^2\right)/C_\psi\left((ap)^2\right) \, .
\label{running}
\ee
The quantity $C_\psi$ is normalised so that $C_\psi(1)=1$.
As can be seen, in
this case the renormalisation group running is coming from $Z_\psi$ alone. It is small,
but it actually improves the scale independence of the data. 
The remain scale dependence of this data is 
very small and a plausible explanation for this is a
small $(ap)^2$ error. Indeed, when a linear fit of the SI data versus 
$(ap)^2$ is performed, for $0.8 < (ap)^2 < 2.0$, the gradient is $\approx 0.01$.

In the case of $Z_\psi/Z_S$, the two renormalisation constants are both running
with $(ap)^2$. Fig.~\ref{figvssi} shows both $\Gamma_S$ and the scale invariant (SI)
quantity after three loop running. The modification to the raw data, which is taken
after the mass-pole has been subtracted gives a satisfactory result. The linear fit of the SI data 
versus $(ap)^2$ in the range of $0.8 < (ap)^2 < 2.3$, 
gives a gradient of about 0.02. 

The SI result of $\Gamma_T$ which comes from the two loop running of $Z_T$~\cite{Gracey} 
and three loop running of $Z_\psi$ is plotted in Fig.~\ref{figvtsi},
The ``bare'' data is taken by a simple linear extrapolation to the chiral limit.
The linear fit to the SI data gives gradient about 0.01.  
 
With the interpretation that the remaining scale dependence is due
to ${\mathcal O{(ap)^2}}$ effects, the correct way to extract the renormalisation
coefficients is to first construct the SI quantity as described above, and
then fit any remaining scale dependence~\cite{noi_mq} to
the form
\be
y = c_1 + c_2  (ap)^2 \, ,
\label{sifit}
\ee
for a range of momenta that is chosen to be ``above'' the region for
which condensate effects are important.  We choose the range $ 0.8 < (ap)^2 < 2.3$.
We then  apply the renormalisation group running formula (an inverse operation of
Eq.~(\ref{running})~) to $c_1$ in Eq.~(\ref{sifit}) to
get the ratio of the renormalisation constant which is now free of $(ap)^2$ error in
RI scheme.

\subsection{ Extracting $Z_\psi$ from the propagator}

  Here we will use Eq. (\ref{eq:Z_q'_WI}) to extract
the wave function renormalisation constant $Z_\psi$. We will use two
different definitions of lattice momenta, {\it i.e. }, the discrete lattice momentum
defined in Eq.~(\ref{eq:pi}) and the kinematic lattice momentum $q_\mu$ which was introduced in
Eq. (A9) of Ref.~\cite{overlgp}.

 
\subsubsection{Discrete lattice momentum $p_\mu$}

\begin{figure}[t]
{\small
\resizebox*{\columnwidth}{4.8cm}{\rotatebox{90}{\includegraphics{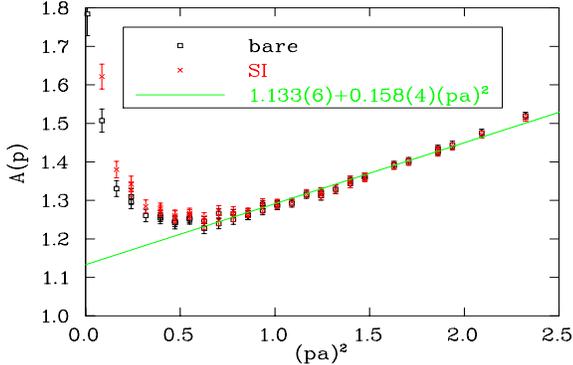}}}
}
\vspace*{-1.2cm}
\caption{\label{zpsifp}{\small
$A(p)$ in the chiral limit and its scale invariant counterpart versus $(pa)^2$ as $ a~+~b~(pa)^2 $.
The fitting range is [0.70, 2.32], which gives $Z_\psi$  about $1.133 \pm 0.006$.
}}
\vspace*{-0.8cm}
\end{figure}

We here use the discrete lattice momentum $p_\mu$  in Eq.~(\ref{eq:Z_q'_WI})
to calculate $Z'_\psi$.
By using  Eq.~(\ref{eq:Z_q'_WI}), we can get the
corresponding quantity of $Z'_\psi$ (here we use 
the symbol $A(p)$ for $Z'_\psi$).
The resulting $A(p)$ is strongly dependent on the scale. We use three
loop perturbative running in RI scheme and finally  
extract the value of $Z_\psi$ from $A(p)$ by fitting $A(p)$ to \Eq{sifit} 
for a range of momenta for 
which the condensate effects are unimportant. Here we chose
the range $(ap)^2~ \subset (0.70, 2.32)$, which corresponds to $p \in$ (1.78, 3.23) GeV.
$c_1$ can be identified as the SI value of $Z_\psi$.
In Fig.~\ref{zpsifp} we show a linear fit for $A(p)$ versus $(pa)^2$
as in Eq.~(\ref{sifit}).

\subsubsection{Kinematic lattice momentum $q_\mu$}

\begin{figure}[t]
{\small
\resizebox*{\columnwidth}{4.8cm}{\rotatebox{90}{\includegraphics{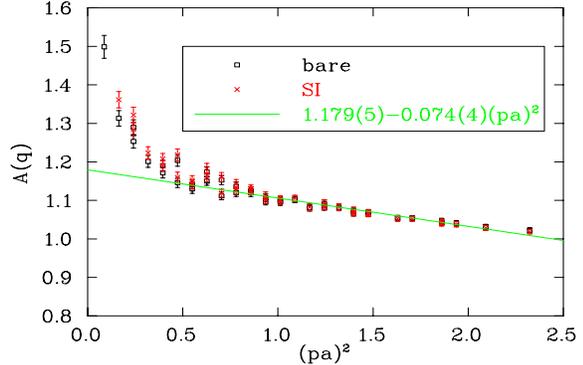}}}
}
\vspace*{-1.2cm}
\caption{\label{zpsifq}{\small
$Z_\psi$ calculated from \Eq{eq:Z_q_WI} by using the kinematic lattice momentum $q_\mu$
instead of the  discrete lattice momentum $p_\mu$.
}}
\vspace*{-0.8cm}
\end{figure}
 Another way to calculate $Z_\psi$ is to use the kinematic lattice momentum $q_\mu$
defined in Eq. (A9) of Ref.~\cite{overlgp}. The calculations are similar to those
using $p_\mu$. The result is plotted in \Fig{zpsifq}. The fitted value of $Z_\psi$
is 1.179(5), with the difference to the value calculated by $p_\mu$ being less than 4\%.
In both calculations, we attribute the remaining $(pa)^2$ dependence 
to the possible ${\mathcal O((pa)^2)}$  error and use the linear fit Eq.(~\ref{sifit}) to
remove it. When using the kinematic lattice momentum $q_\mu$
to calculate $Z_\psi$, the ${\mathcal O((pa)^2)}$ errors are much smaller than in the case of using 
the discrete lattice momentum $p_\mu$. 

\subsection{Matching to $ {\overline{MS}} $ scheme }

In order to confront experiment, one frequently prefers to quote the final
results in the $\overline{MS}$ scheme at a certain scale. For light hadrons,
the popular scale is 2 GeV.

    The perturbative expansion of the ratio $Z^{\overline{MS}}/Z^{RI}$ to two-loop order is given
by~\cite{noi_mq}
\begin{equation}   \label{eq:R}
R = \frac{Z^{\overline{MS}}}{Z^{RI}} = 1 + \frac{\alpha_s}{4 \pi} (Z^{RI})_0^{(1)}
+ (\frac{\alpha_s}{4 \pi})^2 (Z^{RI})_0^{(2)}+... \, .
\end{equation}
The numerical values of the
matching coefficients, $Z^{(1)}_0$ and $Z^{(2)}_0$ in Eq.~(\ref{eq:R})
for $Z_{\psi}$ and $Z_S$ can be found in Appendix C of  Ref.~\cite{Tblum1},
we have not found the numerical value for $Z_T$ in the literature.
The final results for the renormalisation constants are listed in Table~\ref{zresult}.

\vspace*{-0.5cm}
\begin{table}[ht]
\caption{\label{zresult} Final Z-factor results }
\begin{tabular}{ccc}
\hline

$Z-{\rm factor}$  & RI (2 GeV) & $ {\overline{MS}} $ at 2 GeV       \\
\hline
$Z_A/Z_\psi$ & 0.924$\pm$0.004  & 0.928$\pm$0.004    \\
$Z_S/Z_\psi$ & 0.739$\pm$0.003  & 0.842$\pm$0.004    \\
$Z_T/Z_\psi$ & 1.009$\pm$0.002  & 1.012$\pm$0.002    \\
$Z_\psi(from~~p)    $ & 1.134$\pm$0.006  & 1.130$\pm$0.006    \\
$Z_\psi(from~~q)    $ & 1.180$\pm$0.005  & 1.175$\pm$0.005    \\
\hline
\end{tabular}
\end{table}
\vspace*{-0.8cm}

\section{SUMMARY}
\label{sec:conclusions}

In this work  we performed the non-perturbative renormalisation of 
composite operators with overlap fermions.
The main results are: in ${\overline{MS}}$ scheme at $\mu$ = 2 GeV, 
$Z_\psi$ = 1.153(5)(22), $ Z_S $ = 0.970(4)(20),
$ Z_A $ = 1.070(5)(21), and $ Z_T$ = 1.167(3)(23), 
where the first error is statistical
and the second error comes from using the two different lattice 
momenta to get $Z_\psi$.

\vspace*{0.5cm}
Support for this research from the Australian Research Council
is gratefully acknowledged.

\end{document}